\documentclass[twocolumn,showpacs,preprintnumbers,amsmath,amssymb]{revtex4}

\usepackage{graphicx}
\usepackage{dcolumn}
\usepackage{bm}
\usepackage{color}
\usepackage{graphicx}
\usepackage{epstopdf}
\usepackage[hang,scriptsize]{subfigure}

\begin{document}


\title{Epidemic Spreading in Weighted Networks: An Edge-Based Mean-Field Solution}

\author{Zimo Yang}
\author{Tao Zhou}
\email{zhutou@ustc.edu}
\affiliation{Web Sciences Center, University of Electronic Science and Technology of China, Chengdu 610054, People's Republic of China}

\begin{abstract}
Weight distribution largely impacts the epidemic spreading taking place on top of networks. This paper studies a susceptible-infected-susceptible model on regular random networks with different kinds of weight distributions. Simulation results show that the more homogeneous weight distribution leads to higher epidemic prevalence, which, unfortunately, could not be captured by the traditional mean-field approximation. This paper gives an edge-based mean-field solution for general weight distribution, which can quantitatively reproduce the simulation results. This method could find its applications in characterizing the non-equilibrium steady states of dynamical processes on weighted networks.
\end{abstract}


\maketitle

\section{Introduction}

Many social, biological and communication systems can be described as networks with nodes denoting individuals and edges representing interactions among them. Edges in many networks do not only indicate the existence of interactions, but also have associated with weights to record the interacting strengths. Examples are numerous, including the scientific collaboration networks where the edge weight between two scientists depends on the number of co-authorized papers and the number of authors in each of these papers \cite{Newman2001}, the airport networks with edge weight representing the number of flights \cite{Li2004} or the number of available seats \cite{Barrat2004} between two airports, the food webs where edge weight can be defined by interaction frequency, by carbon flow or by interaction strength \cite{Krause2003}, the cell-phone communication networks with weight being assigned as the total number of calls or the total duration of calls \cite{Onnela2007}, the world trade web where the edge weight accounts for the exports over GDP of the exporting country \cite{Fagiolo2008}, and so on.

Weight distribution has remarkable effects on the properties of dynamical processes taking place on top of networks, such as synchronization \cite{Zhou2006a,Zhou2006b,Zhao2006}, transportation and routing \cite{Thadakamalla2005,Wu2006,Ramasco2007}, evolutionary game \cite{Du2008,Cao2011}, cascading failure \cite{Wang2008,Mirzasoleiman2011}, diffusion \cite{Ou2007,Baronchelli2010}, control \cite{Liu2007}, condensation \cite{Tang2006}, and so forth. Epidemic spreading in weighted networks has also been considered in the literature \cite{Zhou2006c}. Yan \emph{et al.} \cite{Yan2005} studied a susceptible-infected (SI) model on weighted scale-free networks and found that the heterogeneous weight distribution will slow down the spreading. Therefore, given the topology and mean infectivity, the epidemic spreads fastest in unweighted networks. Chu \emph{et al.} \cite{Chu2009} investigated the same dynamical process upon weighted scale-free networks with community structure, and showed that the weights of edges among different communities have higher impacts than those of edges inside communities. Chu \emph{et al.} \cite{Chu2011} further studied the susceptible-infected-recovered (SIR) model and found that the weight distribution has considerable impacts on both epidemic threshold and prevalence. Karsai \emph{et al.} \cite{Karsai2006} considered the contact process in weighted scale-free networks, in which the weight of an edge connecting two higher-degree nodes is relatively small. Yang \emph{et al.} \cite{Yang2008} further proved that in the contact process, the epidemic prevalence can be maximized by setting the egde weight inversely proportional to the degree of receiving node.

Mean-field approximation is a powerful tool to analyze dynamical processes on networks. In particular, the so-called heterogeneous mean-field approximation has achieved great success in predicting threshold of epidemic spreading in scale-free networks \cite{Pastor2001a,Pastor2001b}. However, this method may fail to capture the behavior of some dynamical processes in the thermodynamic limit \cite{Castellano2010} and can not account for the fluctuations caused by quenched structure \cite{Dorogovtsev2008}. In this paper, we study epidemic spreading in homogeneous networks with heterogeneous weight distribution, and develop an edge-based mean-field approximation. Simulation results indicate that the more homogeneous the weight distribution is, the higher the epidemic prevalence is. This phenomenon can be quantitatively reproduced by the edge-based mean-field solution. In contrast, the node-based mean-field solution is insensitive to the change of weight distribution, and it tends to overestimate the epidemic prevalence.

This paper is organized as follows. In the next section, the model will be introduced, including how to construct networks, how to generate weight distributions, as well as the details of spreading dynamics. In Section 3, the edge-based mean-field solution will be presented, while for comparison, the node-based mean-field solution will be introduced in the Appendix. Section 4 compares the simulation results and analytical solutions and give an explanation about why the edge-based mean-field solution is more accurate than the node-based one. Eventually, we draw the conclusion, emphasize the advantages of the present method and discuss its potential relevance and applications in the last section.

\section{Model}

To emphasize the effects caused by the weight distribution, we concentrate on the regular random networks where edges are added in a completely random matter and the degree of every node is fixed as $k$. In this way, the impacts of local clustering and heterogeneity in network topology can be eliminated.

We investigate two sets of weight distributions. In the former case, the weight of each edge is the average of $n$ elements independently drawn from the uniform distribution $\mathbf{U}(0,2)$. When $n=1$, the weight distribution $p(w)$ is exactly the same as $\mathbf{U}(0,2)$, and $p(w)$ will become more homogeneous with the increasing of $n$, following the form
\begin{equation}
\begin{aligned}
&p(w;1)=\mathbf{U}(0,2), \\
&p(w;n)=\int^2_0zp(\frac{nw-z}{n-1};n-1)dz, n=2,3,..., \\
&\lim_{n\rightarrow \infty}p(w;n)\sim \mathbf{N}(\mu,\frac{\sigma^{2}}{n}),
\end{aligned}
\end{equation}
where $\mathbf{N}$ stands for the normal distribution, and $\mu=1$ and $\sigma=\sqrt{\int^{1}_{0}x^{2}dx} \approx 0.57735$ are the mean value and standard deviation of the uniform distribution $\mathbf{U}(0,2)$, respectively. Figure 1(a) illustrates example distributions for $n=1,4,7$ and 10.

In the latter case, the edge weight distribution follows a power law
\begin{equation}
p(w)\sim w^{-\gamma}.
\end{equation}
The power-law distributions are generated according to the method in Refs. \cite{Clauset2009,Yang2011} with weights between $w_{\min}=1$ and $w_{\max}=10^4$. Figure 1(b) illustrates example distributions with $\gamma=2$, 3 and 4.

\begin{figure}
\includegraphics[width=3.5in]{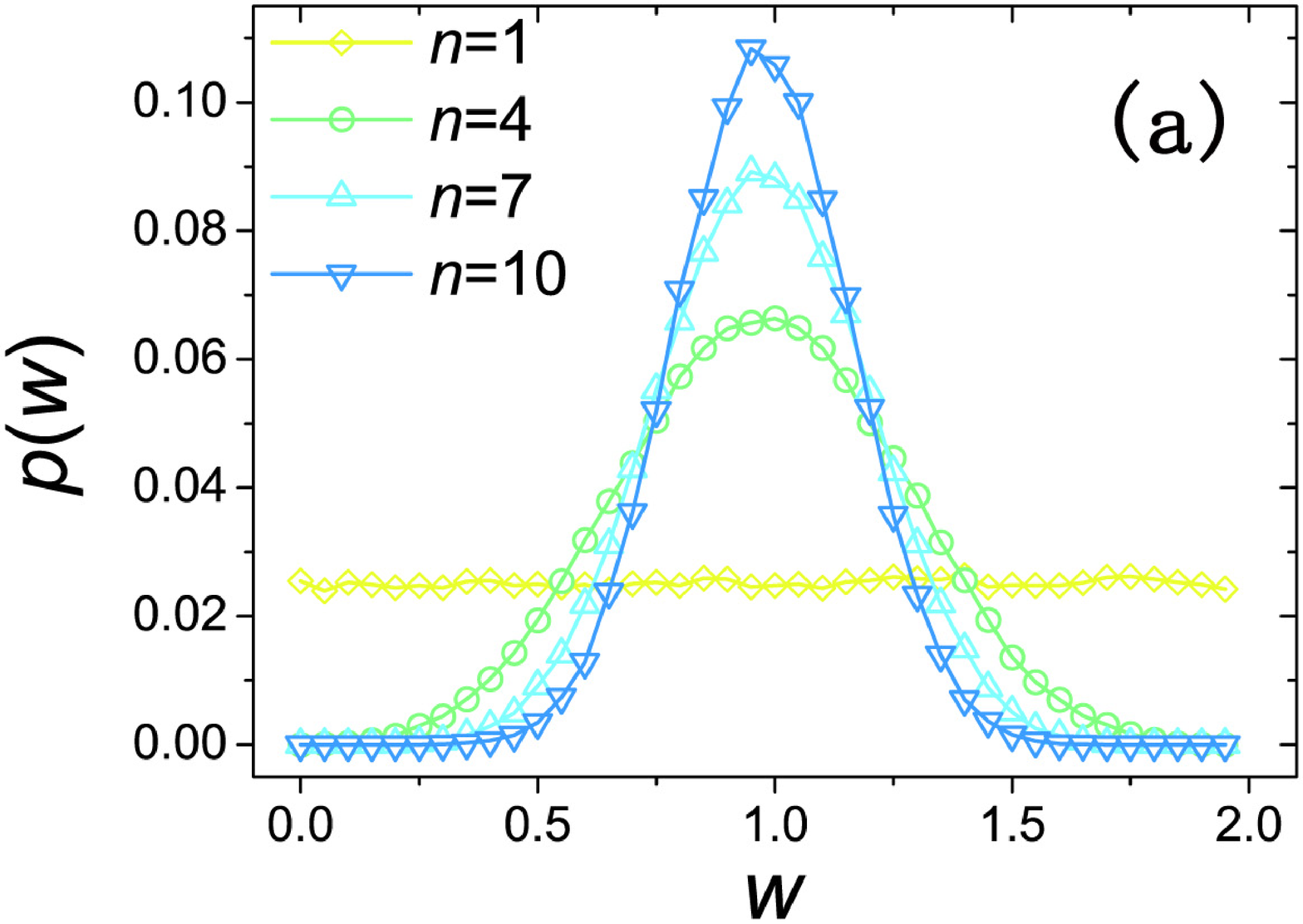}
\includegraphics[width=3.5in]{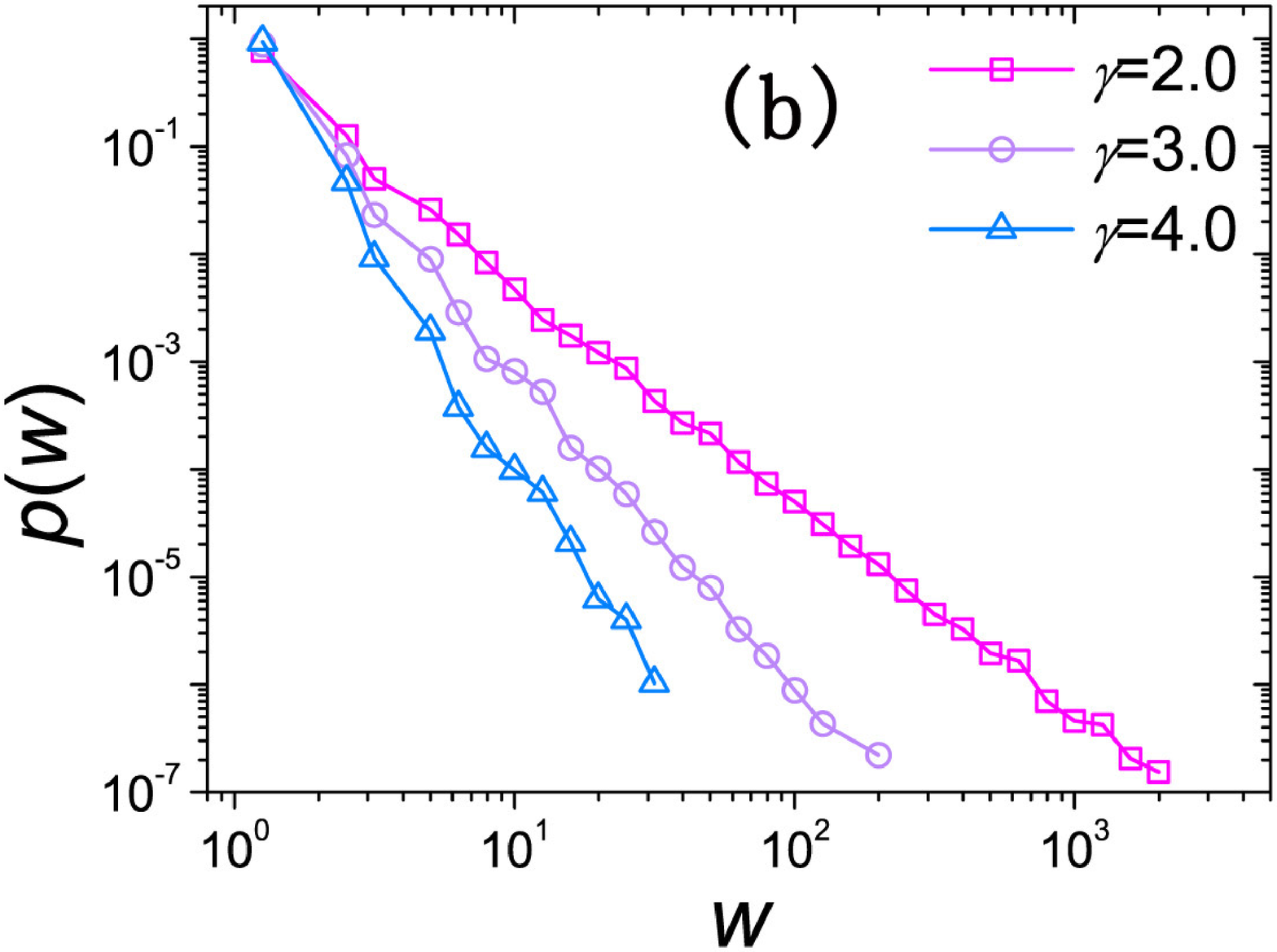}
\caption{(Color online) Illustration of weight distributions: (a) more homogeneous distributions, as described in Eq.~1 and (b) more heterogeneous distributions, as described in Eq.~2. Each distribution is obtained by sampling $10^6$ weights.}
\end{figure}

The epidemic spreading is described by a weighted susceptible-infected-susceptible (SIS) model. Each node is in one of two discrete states, either susceptible (S) or infected (I). At each time step, each infected node will select one of its neighbors to contact with. The selection probability is proportional to the edge weight. If the selected neighbor has been infected already, nothing happens, while if it is susceptible, with probability $\lambda$, it will be infected and then turn to be I-node in the next time step. Meanwhile, each infected node will become susceptible in the next time step with probability $\beta$. We assume that each infected node can contact only one neighbor per time step since in real social networks even an individual may have many friends, her could only interact with a very limited number of them whatever by physical contacts or by information communication \cite{Zhou2006d,Yang2007}.

\section{Analysis}

 The analytical results can be obtained by the Markov chain equations about the evolution of edge states. For every edge in the network, there are three possible states depending on its two ends, say SS, SI and II. With an edge-based mean-field hypothesis, we assume that edges with the same weight distribute uniformly in the network. Given a specific weight $w$, we denote $e_{SS}(w)$, $e_{SI}(w)$ and $e_{II}(w)$ as the densities of edges in corresponding states, which satisfy the normalized condition
\begin{equation}
e_{SS}(w)+e_{SI}(w)+e_{II}(w)=1.
\end{equation}
 In the steady state, not only the density of the infected node, but also the densities of edges in different states keep stable. Denote $\overrightarrow{E}(t,w)=[e_{SS}(t,w),e_{SI}(t,w),e_{SI}(t,w)]^{T}$ the densities of edges in different states at time step $t$. A time-dependent transferring matrix, $T(t,w)$, governs how $\overrightarrow{E}(t,w)$ changes to $\overrightarrow{E}(t+1,w)$ during the time step $t$, so the Markov chain equations can be written as,
\begin{equation}
\overrightarrow{E}(t+1,w)=T(t,w)\overrightarrow{E}(t,w).
\end{equation}

An infected node, no matter it associates with which kinds of edges, will change to a susceptible one with probability $\beta$. Clearly, the more infected nodes a susceptible node connects, the higher probability it will be infected. However, we could not precisely know each susceptible node's local surrounding, so we only consider two different prior conditions: a susceptible node is one end of an SI-edge or it is one end of an SS-edge. We denote $\varphi(t)$ the probability that a susceptible node connecting to an SS-edge with weight $w$ will become an infected node during the time step $t$, then for a susceptible node connecting to an SI-edge with weight $w$, the probability it will be infected during the time step $t$ is $1-[1-\varphi(t)]\left[1-\frac{w}{(k-1)\langle w\rangle+w}\lambda\right]$, where $\langle w\rangle$ is the mean weight. The item $\left[1-\frac{w}{(k-1)\langle w\rangle+w}\lambda\right]$ is the probability that the monitored susceptible node will not be infected by the known infected neighbor with $\frac{w}{(k-1)\langle w\rangle+w}$ denoting the probability this infected neighbor will choose the monitored node to contact with, meanwhile the item $[1-\varphi(t)]$ is the probability that the monitored node will not be infected by the other $k-1$ neighbors. Notice that, the item $\frac{w}{(k-1)\langle w\rangle+w}$ is an approximation only valid for homogeneous weight distribution, which is good enough for the case shown in figure 1(a) yet not quantitatively suitable for the case shown in figure 1(b), which will be discussed later.

Denoting $A$, $B$ and $C$ the probabilities that during the time step $t$, an infected node changes to a susceptible one, an susceptible node as one end of an SS-edge weighted $w$ changes to an infected one and an susceptible node as one end of an SI-edge weighted $w$ changes to an infected one, respectively, then $B$ and $C$ depends on $t$ and $w$. Without lose of generality, we always adopt a normalized weight distribution with $\langle w\rangle=1$, and the changing probabilities can be summarized in the follows.
\begin{equation}
\begin{aligned}
&A=\beta, \\
&B=\varphi(t), \\
&C=1-[1-\varphi(t)]\left(1-\frac{w}{k-1+w}\lambda\right).
\end{aligned}
\end{equation}

\begin{figure}
\includegraphics[width=3.0in]{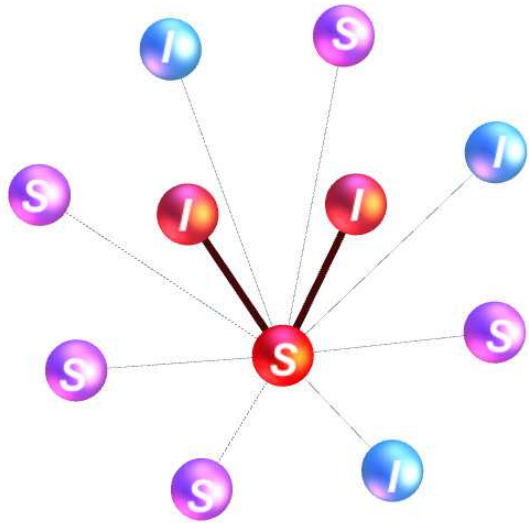}
\caption{(Color online) Illustrations of the densities of SI-edges and active SI-edges. The central node is a susceptible node with 10 neighbors, among which, five are susceptible nodes and the other five are infected nodes. Therefore, the density of SI-edges for the central node is 0.5, and $\rho_{SI}$ is the average density of SI-edges over all susceptible nodes. Among these five infected neighbors, only two of them choose the central node to contact (indicated by red color and thick lines), therefore the density of active SI-edges is 0.2. Analogously, $\rho$ is the average density of active SI-edges over all susceptible nodes.}
\end{figure}

We neglect the indices $t$ and $w$ of $B$ and $C$ for simplicity. Accordingly, the transferring matrix $T(t,w)$ for the changes of densities of edges in different states is
\begin{equation}
T=
\left[
\begin{array}{ccc}
(1-B)^{2} & A(1-C) & A^{2}\\
2(1-B)B & 1-A-C+2AC & 2A(1-A)\\
B^{2} & (1-A)C & (1-A)^{2}\\
\end{array}
\right].
\end{equation}
As a transferring matrix, $T$ is column normalized.

\begin{figure}
\includegraphics[width=3.5in]{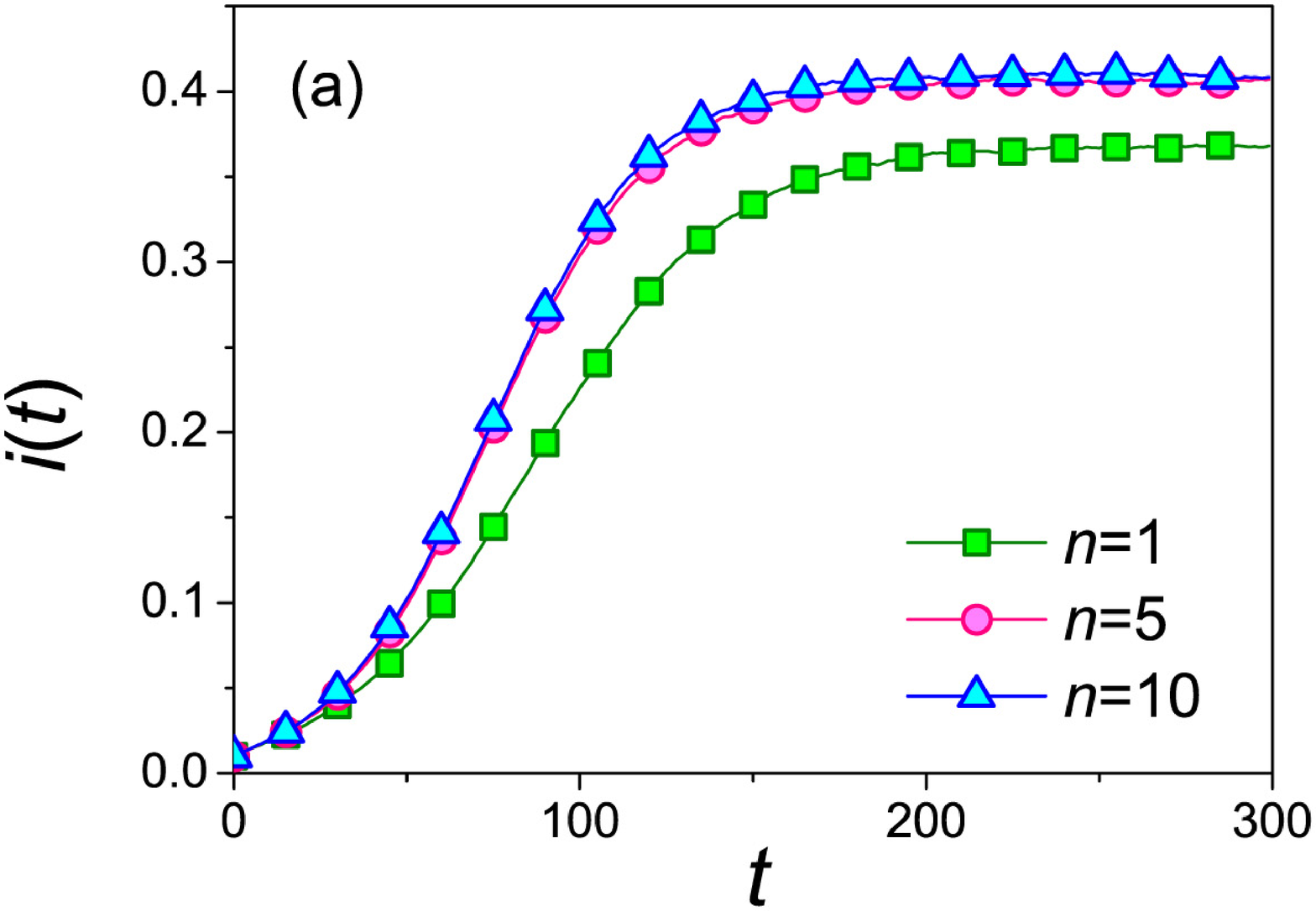}
\includegraphics[width=3.5in]{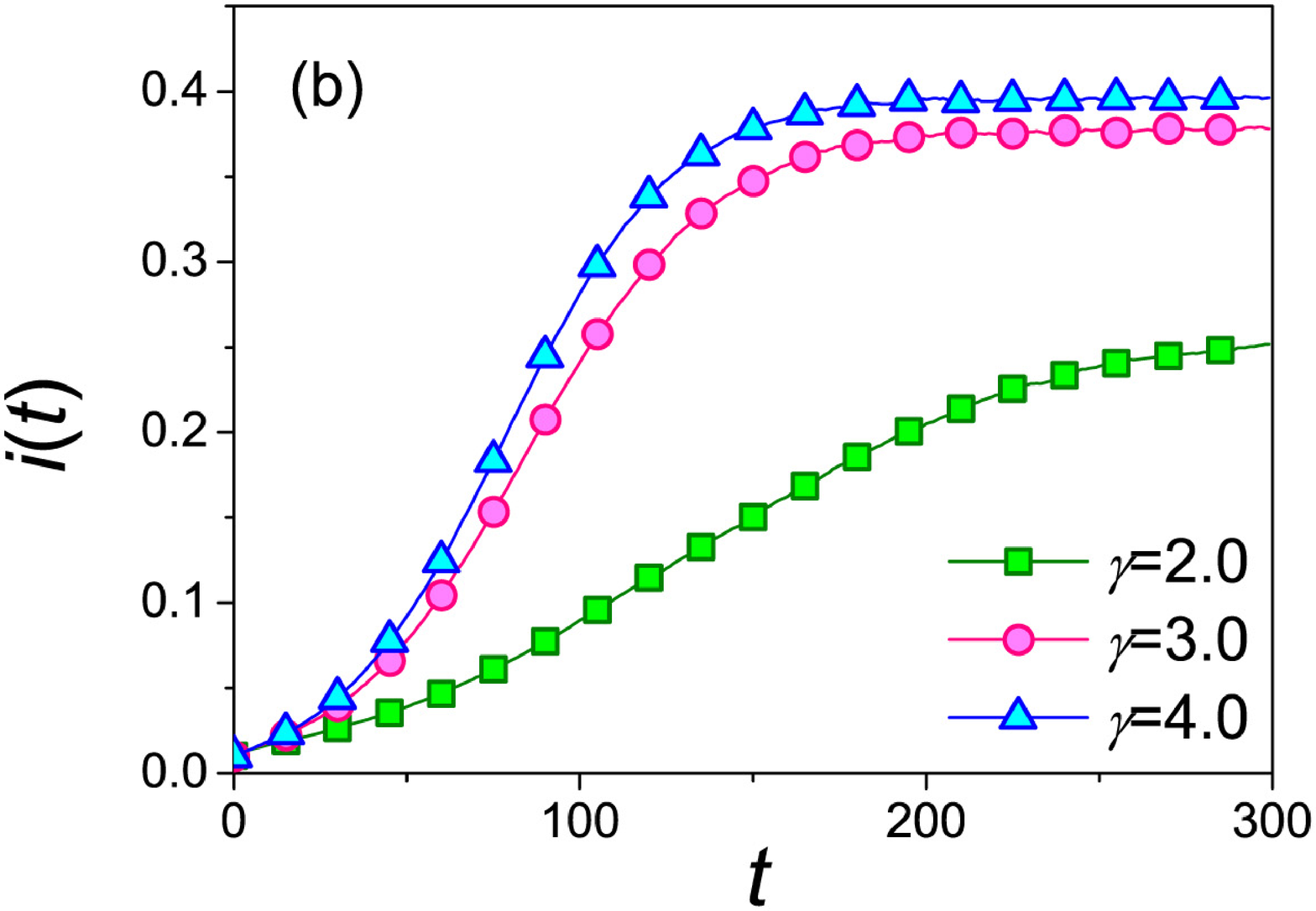}
\caption{(Color online) Density of infected nodes versus time in regular random networks with $N=10^{3}$ and $k=5$. Initially, 10 nodes are randomly selected to be infected, and the infecting rate and recovering rate are $\lambda=0.2$ and $\beta=0.1$, respectively. Every data point is averaged over $10^{3}$ independent realizations. Subgraphs (a) and (b) imply the spreading processes in the networks with different kinds of weight distributions. Three curves in subgraph (a) correspond to Eq.~1 with parameter $n=1$, 5 and 10, while those in subgraph (b) to Eq.~2 with parameter $\gamma=2.0$, 3.0 and 4.0.
}
\end{figure}

Under the above framework, the estimation of $\varphi(t)$ becomes the most significant part. Each susceptible node is associated with $k$ edges, which can only be in SI or SS states. For example, as shown in figure 2, the susceptible node in the center is surrounded by $k=10$ nodes, and the density of its associated SI-edges is 0.5. Denote by $\rho_{SI}$ the average density of SI-edges associated with a randomly selected susceptible node, if the edges in different states distribute uniformly in the network, the number of SI-edges and SS-edges are $M_{SI}=N_Sk\rho_{SI}$ and $M_{SS}=\frac{1}{2}N_Sk(1-\rho_{SI})$, where $N_S$ denotes the number of S-nodes and the factor $1/2$ accounting for the duplicate count of SS-edges. Therefore, the density $\rho_{SI}$ can be written as
\begin{equation}
\rho_{SI}=\frac{M_{SI}}{2M_{SS}+M_{SI}}.
\end{equation}
This simple derivation could find its application in estimating $\varphi(t)$ by further considering (i) edges with different weights should be treated separately to ensure the accuracy, and (ii) an SI-edge is active only if the corresponding $I$-node chooses this edge to make a contact. For example, in figure 2, the density of SI-edges subject to the central node is only 0.2. Accordingly, the average density of active SI-edges associated with a randomly selected susceptible node is
\begin{equation}
\rho(t)=\frac{\int e_{SI}(t,w)p(w)\frac{w}{k-1+w}dw}{\int [e_{SI}(t,w)+2e_{SS}(t,w)]p(w)dw},
\end{equation}
where $\frac{w}{k-1+w}$ is the approximate probability an SI-edge with weight $w$ gets active with homogeneous weight distribution. If there are $l$ active SI-edges connecting to a susceptible node, it will be infected with probability $1-(1-\lambda)^{l}$. With the density of active SI-edges being $\rho(t)$ and the degree of each node being $k$, the infected rate of a susceptible node in an SS-edge is
\begin{equation}
\varphi(t)=\Sigma_{l=1}^{k-1}C_{k-1}^{l}\rho^l(t)[1-\rho(t)]^{k-1-l}\left[1-(1-\lambda)^{l}\right].
\end{equation}

By iteratively multiplied by $T(t,w)$, $\overrightarrow{E}(t,w)$ will rapidly converge to a steady vector $\overrightarrow{E}(w)=[e_{SS}(w),e_{SI}(w),e_{II}(w)]^{T}$, and the density of infected nodes in the non-equilibrium steady state can thus be obtained as
\begin{equation}
i=\frac{1}{2} \int [2e_{II}(w)+e_{SI}(w)]p(w)dw.
\end{equation}

If the weight distribution $p(w)$ is highly heterogeneous like those shown in Fig. 1(b), the estimation of the probability an infected node will choose an edge with weight $w$, $\eta=\frac{w}{k-1+w}$, is not accurate enough. In this case, we replace it by
\begin{equation}
\eta=\int\frac{w}{\phi+w}q(\phi)d\phi,
\end{equation}
where $\phi$ denotes the sum of $k-1$ randomly selected weights from the distribution $p(w)$ and $q(\phi)$ is the distribution of $\phi$. Generally speaking, given $p(w)$, $q(\phi)$ can be easily determined numerically. Therefore, via replacing the term $\frac{w}{k-1+w}$ in Eq. 5 and Eq. 8 by Eq. 11, we can more accurately characterize the systems with heterogeneous weight distributions.

\section{Results}

\begin{figure}
\includegraphics[width=3.5in]{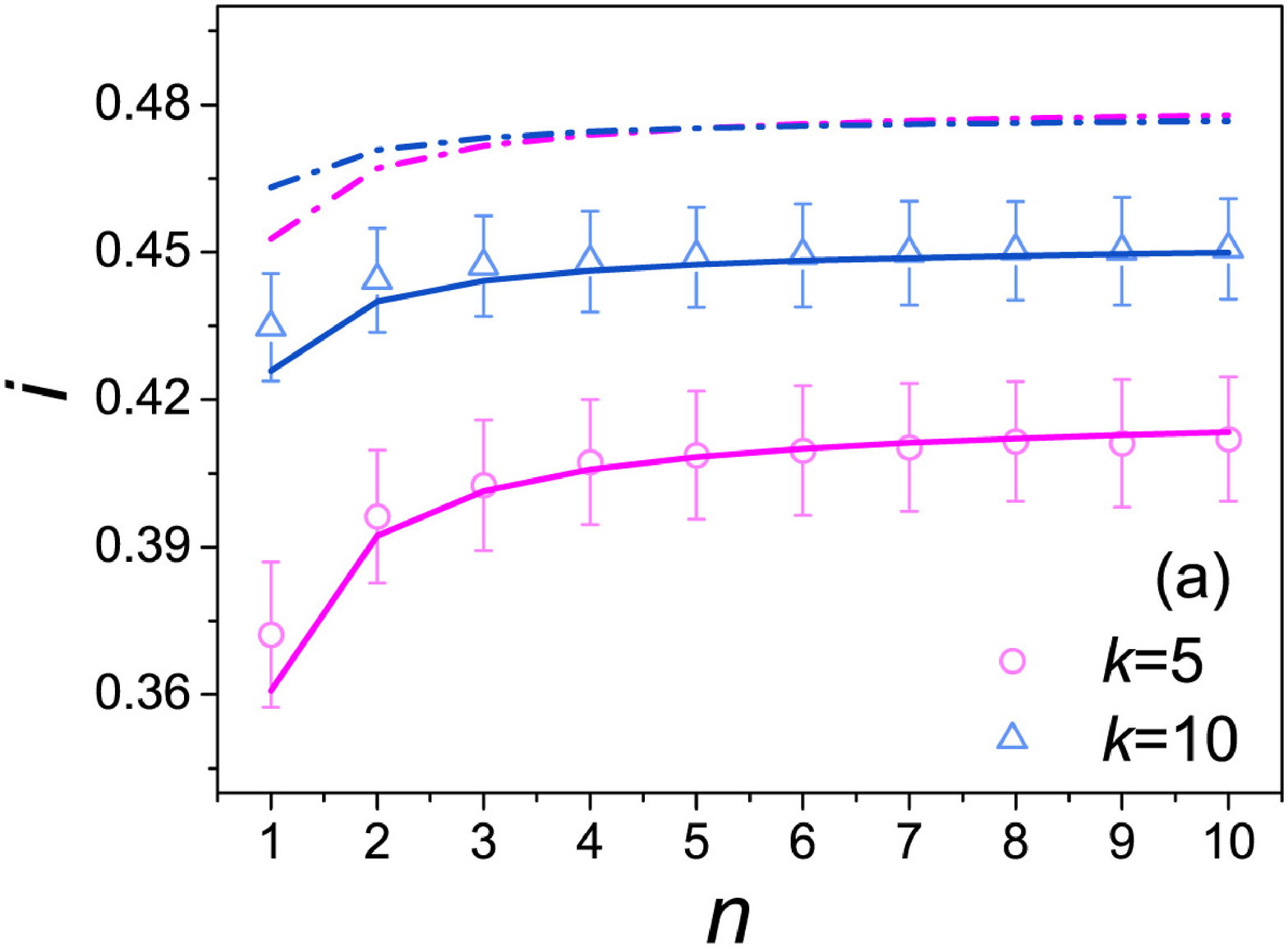}
\includegraphics[width=3.5in]{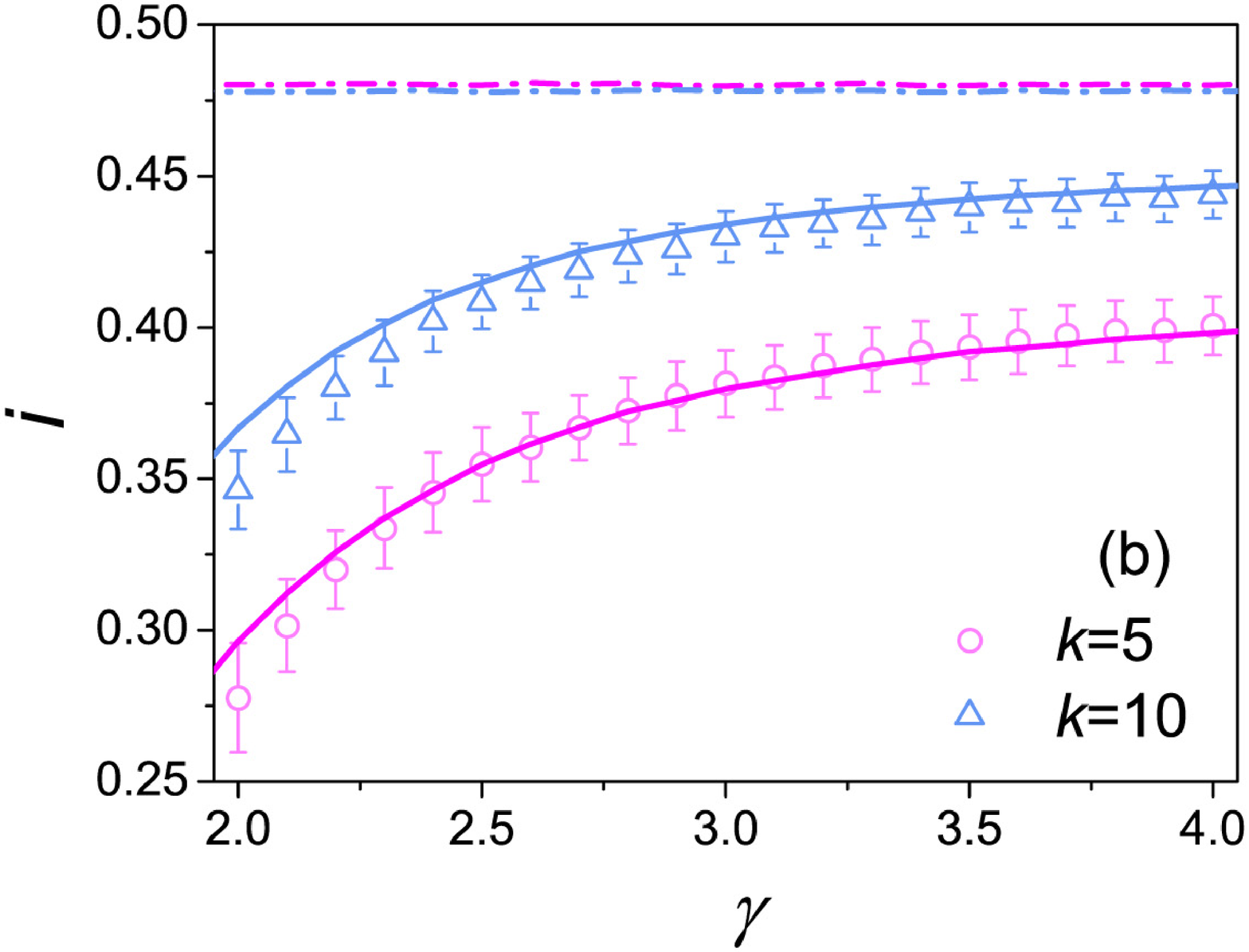}
\caption{(Color online) How the density of infected nodes in the stable state changes with the weight distribution with $N=10^{3}$, $\lambda=0.2$ and $\beta=0.1$ fixed. The red and blue colors represent the cases $k=5$ and $k=10$, respectively. Simulation results (triangles and circles) are the average over $10^{3}$ independent realizations, with the error bars denoting the standard deviations. The solid and dashed lines correspond to edge-based and node-based mean-field solutions, respectively. Similar to figure 2, subgraphs (a) and (b) are for different kinds of weight distributions.}
\end{figure}

We have checked that the steady state is insensitive to the initial condition and the network size, and thus initially we always randomly select 10 nodes as infected and all the shown results are implemented on a network of $10^3$ nodes and averaged over $10^3$ independent realizations. Figure 3 displays the fundamental graph of an SIS model, say the density of infected nodes, $i(t)$, as a function of time. Note that, $p(w)$ will become more homogeneous as the increasing of $n$ and $\gamma$, and at the same time, the density of infected nodes in the steady state increases. In other words, the more homogeneous weight distribution leads to higher epidemic prevalence.

\begin{figure}
\includegraphics[width=3.5in]{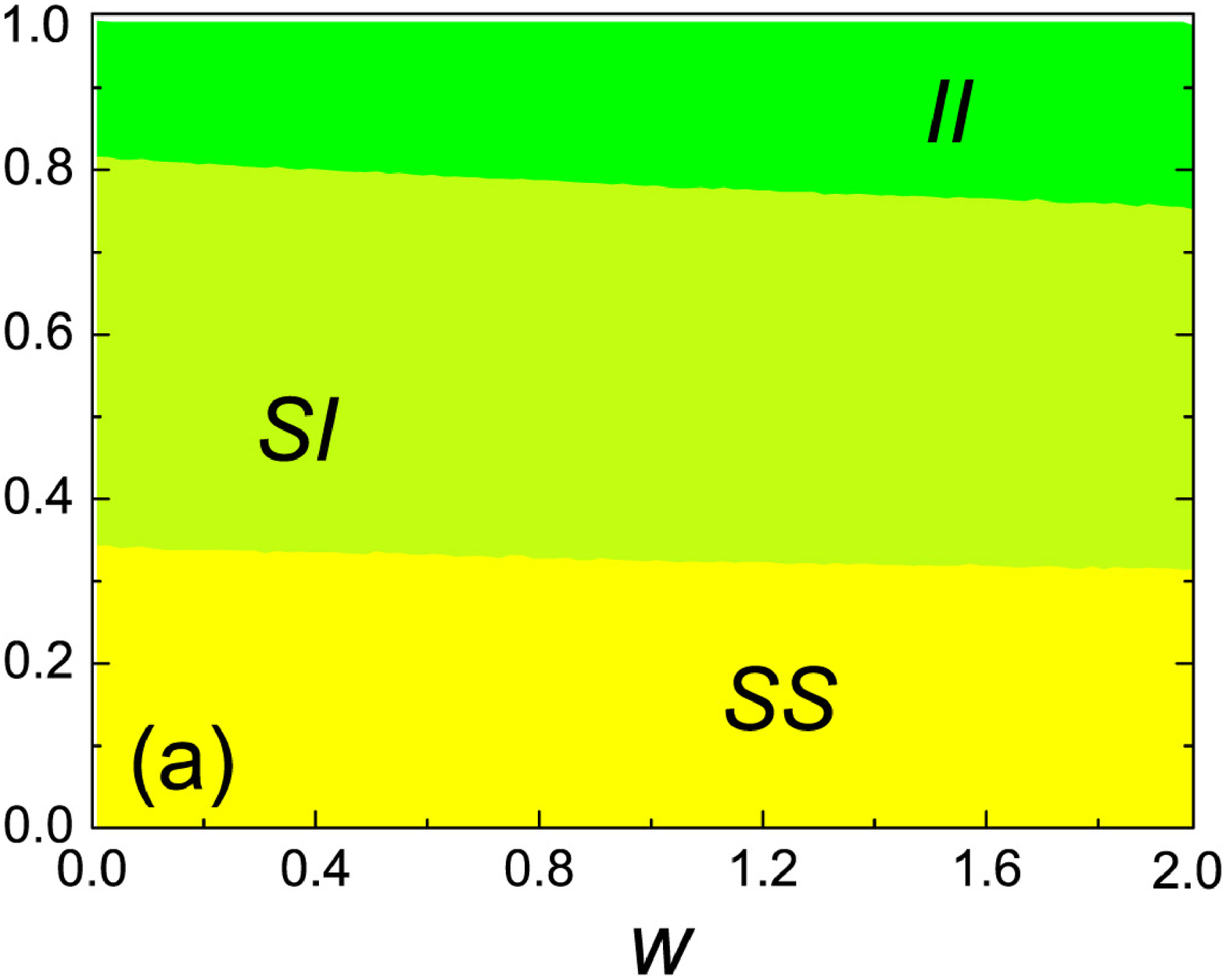}
\includegraphics[width=3.5in]{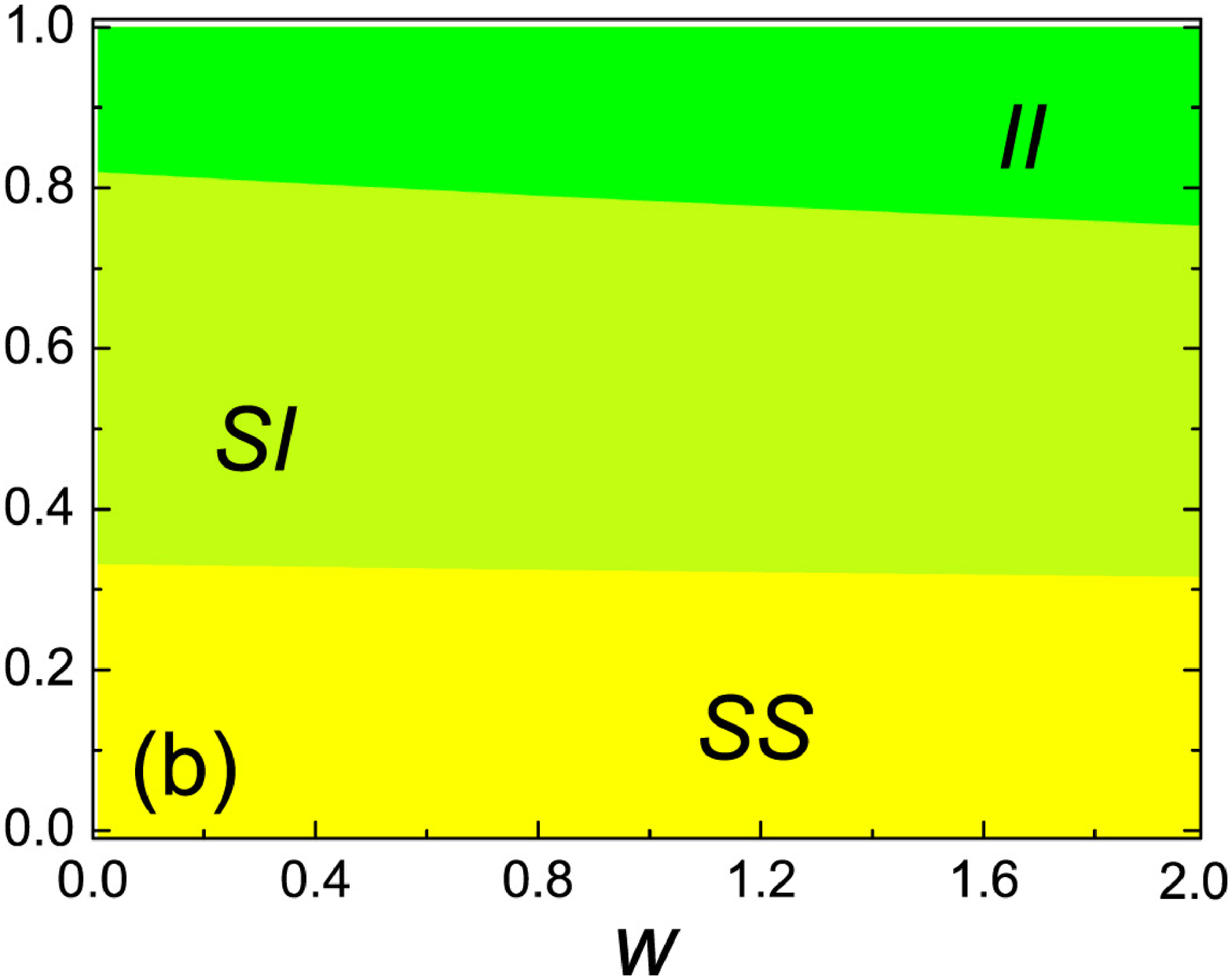}
\caption{(Color online) How the densities of edges in different states (II, SI and SS) change with the weight $w$ in the steady state. The weight distribution is set as $\mathbf{U}(0,2)$ (i.e., $n=1$) and other parameters are $N=10^{3}$, $k=10$, $\lambda=0.2$ and $\beta=0.1$. Subgraphs (a) and (b) are the simulation result (averaged over $10^{3}$ independent realizations) and analytical solution, respectively. }
\end{figure}

To further explore how the weight distribution affects the density of infected nodes in the steady state, $i$, we show in figure 4 how $i$ changes with $n$ and $\gamma$. It is consistent with figure 2 that the larger $n$ or $\gamma$ is, the more infected nodes exist in the stable state. In order to emphasize the advantages of the edge-based mean-field solution, not only do we compare it with simulation results, but with the classical node-based mean-field solution as well, which is introduced in \textbf{Appendix A}. Very clearly, the edge-based mean-field solution can quantitatively reproduce the simulation results, while the node-based mean-field solution completely fail to capture the impacts of weight distribution.

Figure 5 provides an evidence that could, to some extent, explain the advantage of edge-based mean-field approximation. Even for the case $n=1$ corresponding to the most homogeneous weight distribution discussed in this article, the densities of edge states change with the weight $w$. This will become even more remarkable when the weight distribution gets broader. The edge-based mean-field solution (see figure 5b) can quantitatively capture the weight-dependent state densities (see figure 5a). This figure also clearly points out the causation of the failure of node-based mean-field approximation, that is, the ignorance of the different density of edge states at different weights.

\section{Discussion}
To solve the SIS model on weighted regular random networks, this paper proposed an edge-based mean-field method, which outperforms the traditional node-based method and can quantitatively reproduce the simulation results. Both analytical solution and simulation results show that the more homogeneous weight distribution leads to higher epidemic prevalence.

The main shortcoming of the mean-field approximation in network analysis is its disability in capturing the fluctuations caused by quenched structure. Compared with the traditional node-based mean-field method \cite{Pastor2001a,Pastor2001b}, more accurate solutions could possibly be obtained by extending the analyzed units from nodes to edges, form edges to 3-order subgraphs, and so on. The mean-field approximation for correlated networks \cite{Boguna2002}, to some extent, can be considered as an elegant example in this direction since the degrees of two ends of an edge are simultaneously considered. The present work also goes in such direction, but it directly considers the heterogeneity of edge weights instead of that of node degree.

The present method could definitely find applications in analyzing dynamical processes on weighted networks, such as cascading, voting model, evolutionary game, transportation, etc. However, in reality, both node degrees and edge weights can be highly heterogeneous, and neither of the traditional node-based mean-field approximation nor the present edge-based mean-field approximation is able to give very accurate predictions. How to integrate these two methods is still an open issue.

\begin{acknowledgments}
This work is partially supported by the National Natural Science Foundation of China under grant number 90924011, and the Fundamental Research Funds for the Central Universities.
\end{acknowledgments}

\appendix

\section{Node-Based Mean-Field Solution}

Denoting $N_{S}(t)$ the number of susceptible individuals at time step $t$ and $\rho_{SI}(t)$ the average density of SI-edges associated with a randomly chosen susceptible node, the number of SI-edges and SS-edges are
\begin{equation}
\begin{aligned}
&M_{SS}(t)=\frac{1}{2}N_{S}(t)k(1-\rho_{SI}(t)), \\
&M_{SI}(t)=N_{S}(t)k\rho_{SI}(t).
\end{aligned}
\end{equation}
Therefore, the density $\rho_{SI}(t)$ can be written as
\begin{equation}
\rho_{SI}(t)=\frac{M_{SI}(t)}{M_{SI}(t)+2M_{SS}(t)}.
\end{equation}
Denoting $i(t)$ the density of infected individuals, according to the mean-field approximation, $\rho_{SI}(t)=\frac{iN}{N-1}\approx i(t)$ when $N\gg 1$. Similar to the edge-based analysis, the average density of active SI-edges associated with a randomly selected susceptible node is
\begin{equation}
\begin{aligned}
\rho(t)
&=\rho_{SI}(t)\int p(w)  \int\frac{w}{\phi+w}q(\phi)d\phi  dw,\\
&\approx i(t)\int p(w)  \int\frac{w}{\phi+w}q(\phi)d\phi  dw.
\end{aligned}
\end{equation}
Therefore, the infected rate of a susceptible node is
\begin{equation}
\varphi (t)=\Sigma_{l=1}^{k-1}C_{k-1}^{l}\rho(t)^{l}(1-\rho(t))^{k-1-l}[1-(1-\lambda)^{l}].
\end{equation}

Knowing the infecting rate and the recovering rate, $\beta$, the density of infected individuals evolves as
\begin{equation}
i(t+1)=i(t)(1-\beta)+\left[1-i(t)\right]\varphi(t).
\end{equation}
It will rapidly converge to a stable value $i$. The results are shown as dashed lines in figure 3.

\end{document}